\documentstyle[12pt,a4,epsfig]{article}

\begin{document}

\begin{titlepage}
\noindent
\begin{flushright}
PRA -- HEP 96/2\\
March 1996\\
\end{flushright}

\vfill

\begin{center}
\noindent
{\huge\bf{Higgs decay into two photons,
dispersion relations  and trace anomaly}}

\vspace{1cm}

\noindent
{\large Ji\v{r}\'{\i} Ho\v{r}ej\v{s}\'{\i}{\footnote{e-mail: {\tt
Jiri.Horejsi@mff.cuni.cz}}} and Miroslav St\"{o}hr{\footnote{e-mail:
{\tt Miroslav.Stohr@mff.cuni.cz}}} \\}
{\it Nuclear Centre, Faculty of Mathematics and Physics, Charles University\\
V~Hole\v{s}ovi\v{c}k\'{a}ch 2, Prague 8, Czech Republic\\}

\end{center}

\vfill

\begin{center}
{\bf Abstract}
\end{center}

We examine the contribution of $W$ boson loops to the amplitude
of the process $H \longrightarrow \gamma \gamma$ within the
dispersion relation approach, taking up an issue raised in this
context recently. We show that the non-vanishing limit of the
relevant formfactor for $m_{W} \rightarrow 0$ is due to a finite
subtraction induced by the value of the corresponding trace
anomaly. The argument can be turned around and one thus arrives at a
dispersive (``infrared'') derivation of the trace anomaly.

\vspace{1cm}
\\ \hspace*{1.885cm} $W$ boson loops

\vfill

\end{titlepage}

The decay process $H \longrightarrow \gamma \gamma$
occupies an important niche in
the physics of Standard Model (SM), and it has been treated in
the literature in considerable detail (see e.g.\cite{b1} --
\cite{b6} and further
references contained therein). In a recent paper, K\"{o}rner et
al. \cite{b7}
considered the case of large Higgs mass and calculated some two-loop
contributions to this process, in an approximation relying on the
Goldstone-boson (GB) ``equivalence theorem'' for closed $W$ loops
(which presumably should work in the heavy Higgs limit -- cf.
e.g. \cite{b8}),
and using the technique of dispersion relations. Reproducing first the
relevant one-loop results, the authors \cite{b7}
pointed out some peculiar features
of such a dispersive calculation, which remind one of the ``infrared''
aspects of the axial anomaly revealed through the imaginary part of
the famous $VVA$ triangle graph \cite{b9}, \cite{b10}.
To put it in explicit terms,
let us first introduce some notation. Let a (dimensionless) formfactor
$F_W = F_W \left( m_W^2 / m_{H}^{2} \right)$
describing the contribution of $W$
boson loops be defined through
\begin{equation}
\label{eq1}
{\cal M} \left( H \longrightarrow \gamma \gamma \right) =
\frac{\alpha}{2 \pi} \frac{1}{v} F_{W}
\left( k.p \,  g^{\mu \nu} - p^{\mu}
k^{\nu} \right) \varepsilon^{\ast}_{\mu} \left( k~\right)
\varepsilon^{\ast}_{\nu} \left( p \right)
\end{equation}
where $\alpha$ is the fine structure constant, $v=\left( G_{F}
\sqrt{2} \right)^{-1/2}$ is the electroweak mass scale,
and other symbols have an obvious meaning. In what follows we
denote the $W$ mass simply as $m$ for brevity.
The observation made in \cite{b7} consists in the following:
If the $F_W$ is calculated within the ``GB approximation''
by means of unsubtracted
dispersion relation, it does not vanish for $m_H \rightarrow \infty$
(or, equivalently, for $m \rightarrow 0$),
while the corresponding imaginary part vanishes in the naive
limit $m \rightarrow 0$. Thus, in the approach \cite{b7}
one encounters another example of
a situation where the suppression factor proportional to a mass  is
compensated after a pertinent integration;  this is indeed closely
analogous, at least technically, to the dispersive treatment of the
axial anomaly \cite{b9}, \cite{b10} (in particular, a relevant imaginary part
becomes proportional to a $\delta$-function).
The  limit  recovered  in \cite{b7}
within the GB approximation reads
\begin{equation}
\label{eq2}
\lim_{m \rightarrow 0} F_{W} \left( m^{2} / m_{H}^{2} \right) =
2
\end{equation}
which coincides with the value given by the exact one-loop
calculation \cite{b2}.
(Needless to say, when considering the limit $m \rightarrow 0$,
the $v$ is kept fixed.)

In this context, one may remember that the SM Higgs boson is coupled to the
other fields through the trace of the ``improved'' energy-momentum
tensor \cite{b11},
i.e. essentially through the corresponding mass terms (this has been noticed
already in the classic paper \cite{b1}). At quantum level, the trace of the
energy-momentum tensor is known to be anomalous
\cite{b12}, \cite{b13} (i.e. it does
not vanish in the massless limit when quantum corrections are taken into
account) so one may wonder if the non-zero value of the limit
(\ref{eq2}) could
be related to the well - known trace anomaly.

In the present note an attempt is made to elucidate this issue. We examine
the dispersion relation (DR) approach to the relevant Feynman graphs and
show that the limiting value (\ref{eq2})
is determined by a ``sum rule'' for the
imaginary part of the $W$ boson loops together with the value of the
corresponding contribution to the trace anomaly. Later on we will also
comment on the Goldstone boson approximation used in \cite{b7}. The order of
the argument may then be reversed to arrive at a dispersive (``infrared'')
derivation of (the vector boson contribution to) the trace anomaly.
Throughout our discussion we restrict ourselves to one-loop diagrams.

Let us start with an explicit expression for the imaginary part of the
formfactor $F_W$ (defined by eq.(\ref{eq1})),
corresponding to $W$ loops in the unitary
gauge. This can be calculated by means of the Cutkosky rules which give the
discontinuity associated with the standard cut $\left( 4m^2 ,
\infty \right)$  w.r.t.
kinematical variable $t = (k~+ p)^2$ (see fig. 1). One gets
\begin{equation}
\label{eq3}
\mbox{Im}\, F_{W} \left( t; m^{2} \right) = \frac{3 \pi}{2}
\frac{4m^{2}}{t} \left( 2 - \frac{4 m^{2}}{t} \right) \ln
\frac{1+r}{1-r}
\end{equation}

\begin{figure}[hhhh]
\epsfig{file=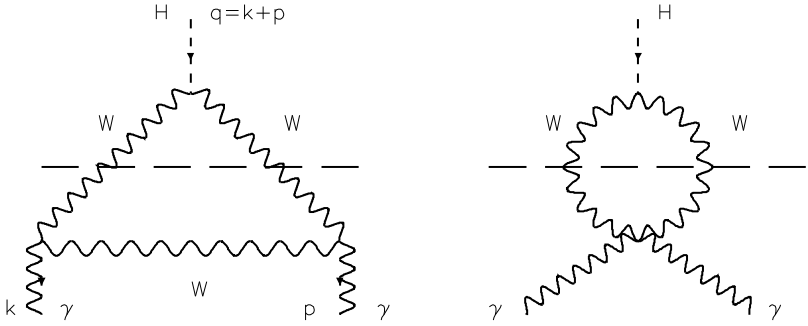}
\protect\caption{{\it Cut diagrams contributing to the imaginary part of
the formfactor $F_W$~given~in}\/ Eq.(3)}
\end{figure}

\noindent
for $t > 4m^2$, where we have denoted $r = \left( 1-
4m^{2}/t \right)^{1/2}$.
Of course, the expression (\ref{eq3}) coincides  with that contained in
the widely quoted result for the $F_W$ (cf. e.g. \cite{b4},
\cite{b5}), obtained
first in \cite{b2} by a direct Feynman graph calculation. We refer here
to the alternative calculation via Cutkosky rules since one would
like to avoid any explicit ultraviolet regularization when adopting
the DR approach. With the imaginary part (\ref{eq3}) at hand, one may try to
evaluate the complete formfactor by means of a dispersion relation.
Let us consider first the unsubtracted form
\begin{equation}
\label{eq4}
F_{W}^{(un)} \left( q^{2}; m^{2} \right) = \frac{1}{\pi}
\int_{4m^{2}}^{\infty} \frac{\mbox{Im}\, F_{W} \left( t; m^{2}
\right)}{t-q^{2}} dt
\end{equation}
for an arbitrary  value of $q^{2} (= m_H^{2})$.
In view of (\ref{eq3}) it is clear that
the integral in the last expression is convergent (reflecting thus the
fact that the total contribution of considered loops is ultraviolet
finite in a renormalizable theory) but one may worry about possible
finite subtractions in the considered DR if one wants to reproduce
the result of the direct Feynman diagram calculation \cite{b2}. It turns
out that such a subtraction is indeed necessary (in contrast with
the scalar-loop method
adopted in \cite{b7}), and its value may be deduced from the
$W$ loop contribution to the trace anomaly. To see this, let us consider
the limit of $F_W (q^2 ; m^2) $  for $q^2 \rightarrow 0$. One gets
\begin{equation}
\label{eq5}
F_{W}^{(un)} \left( 0; m^{2} \right) = \frac{1}{\pi} \int_{4
m^{2}}^{\infty} \mbox{Im}\, F_{W} \left( t; m^{2} \right)
\frac{dt}{t}
\end{equation}
and the integral in the last expression is easily evaluated to
give
\begin{equation}
\label{eq6}
\frac{1}{\pi} \int_{4m^{2}}^{\infty} \mbox{Im}\, F_{W} \left( t ;
m^{2} \right) \frac{dt}{t} = 5
\end{equation}
On the other hand, the value of $F_W ( 0 ; m^{2} )$ is fixed by a low-energy
theorem (see \cite{b1} -- \cite{b4}, \cite{b6}), which leads to the conclusion
that such a quantity is equal to the coefficient $b_W$ of the
$\beta$-function
associated with charge renormalization (due to the charged $W$ bosons in
the present context) defined by
\begin{equation}
\label{eq7}
\beta_{W} \left( e \right) = - b_{W} \frac{1}{16 \pi^{2}} e^{3}
\end{equation}
Let us emphasize that this is precisely the point where the  trace anomaly
enters the picture: In view of the structure of the SM Higgs couplings,
the quantity $F_W (0 ; m^2 )$ actually describes a two-photon matrix element
of the trace of the energy-momentum tensor (cf. \cite{b1}) and this in turn
is determined by the relevant $\beta$ - function (see
\cite{b13}). At one-loop
level one has $b_W  = 7$ (see \cite{b2}
and also the recent paper by Kniehl and
Spira \cite{b6}), so we get a ``boundary condition'' for the DR representation
of the formfactor $F_W$:
\begin{equation}
\label{eq8}
F_{W} \left( 0; m^{2} \right) = 7
\end{equation}
Comparing now the last result with (\ref{eq5}), (\ref{eq6}),
it is seen that the sum
rule (\ref{eq6}) does not saturate the value of the trace anomaly manifested
in (\ref{eq8}); it means that one has to include a finite subtraction in the
DR representation of the $F_W$, namely
\begin{equation}
\label{eq9}
F_{W} \left( q^{2}; m^{2} \right) = F_{W}^{(un)} \left( q^{2};
m^{2} \right) +2
\end{equation}
Using (\ref{eq3}), it is not difficult to prove that
\begin{equation}
\label{eq10}
\lim_{m \rightarrow 0} F_{W}^{(un)} \left( q^{2}; m^{2} \right)
= 0
\end{equation}
From (\ref{eq9}) and (\ref{eq10})  the limit (\ref{eq2})
then follows immediately.

It is worth noticing that one has to make a finite subtraction in the
dispersion relation for $W$ boson loops in the unitary gauge even though
the relevant integral is convergent (for example, no such subtraction
is needed for the contribution of a fermion loop to the same process).
In a sense, the reasoning presented here bears some resemblance with
the dispersive treatment of the axial anomaly: When adopting the
usual tensor basis for the $VVA$ triangle graph \cite{b14}, one also has
to include finite subtraction in a convergent dispersion integral
for the relevant formfactor if the axial anomaly is to be reproduced
correctly (or, in other words, if the vector current is to be conserved).

Let us now return briefly to the observations made in \cite{b7}.
In fact, the
technique of ``GB approximation'' employed in \cite{b7}
enables one to reformulate
the preceding discussion concerning the trace anomaly in the following way.
The contribution of $W$ boson loops may be calculated in a renormalizable
('t Hooft-Feynman) gauge. For $m_H^2 \gg m^2$, the contribution of the
unphysical (Goldstone) scalar counterparts of the $W$  is supposed
to be dominant and the leading contribution to the formfactor $F_W$
is evaluated by means of {\it unsubtracted}\/ dispersion relation
\begin{equation}
\label{eq11}
F_{W}^{(un)} \left( q^{2}; m^{2} \right) = \frac{1}{\pi}
\int_{4m^{2}}^{\infty} \frac{\mbox{Im}\, F_{W}^{(GB)} \left( t;
m^{2} \right)}{t-q^{2}} dt
\end{equation}
where, according to \cite{b7}
\begin{equation}
\label{eq12}
\mbox{Im}\, F_{W}^{(GB)} \left( t; m^{2} \right) = - \pi m_{H}^{2}
\frac{4m^{2}}{t^{2}} \ln \frac{1+r}{1-r}
\end{equation}
with $r$ being defined as in (\ref{eq3}). Note that the factor
of $m_H^2$  in the
last expression comes from the coupling constant for the interaction
of the unphysical Goldstone scalars with physical Higgs boson. (To
avoid confusion, let us also stress that the result (\ref{eq12})
{\it does not}\/
correspond to the expression for ``physical'' scalar loops usually quoted
in the literature --- see e.g. \cite{b4}, \cite{b5} ---
precisely because of  a different
dependence of  the relevant coupling constants on the  masses involved.)
In the limit $m \rightarrow 0$, the expression (\ref{eq12})
becomes  $- 2 \pi m_H^{2} \delta (t)$
(in striking analogy with the calculation of the axial anomaly due
to Dolgov and Zakharov \cite{b9}) and using (\ref{eq11})
one then arrives immediately
at the result (\ref{eq2}).
In such a way, the GB approximation provides a natural
``infrared'' explanation for the finite subtraction in
(\ref{eq9}): This may be
understood as a contribution of unphysical Goldstone bosons (or,
equivalently, of the longitudinal degrees of freedom  associated
with  massive  $W$ bosons). Now, with a simple interpretation of
the subtraction constant in (\ref{eq9}) at hand, one may invoke the sum
rule (\ref{eq6}) to recover eq.(\ref{eq8}).
Thus, from this point of view we have
an alternative derivation of the ``canonical'' value of the $W$ boson
contribution to the trace anomaly in terms of dispersion relations
and ``infrared'' properties of the relevant one-loop graphs. Of course,
such an analysis has one obvious flaw: It is not clear {\it a priori}\/
that the unsubtracted DR in (\ref{eq11}) is the correct form --- in fact it
is an ``educated guess'' which, strictly speaking, is only justified
in confrontation with other results. Let us also remark that the
Goldstone-boson equivalence theorem has been proved rigorously for
the {\it external}\/ vector bosons \cite{b15} (see also \cite{b16}
for a recent reference),
while its applicability for closed loops of massive vector bosons has
only been tested on some specific examples (cf. e.g. \cite{b8}).

In concluding, let us summarize briefly the preceding considerations.
Within the dispersion relation approach to $W$ boson loops (in the unitary
gauge) the zero-mass limit (\ref{eq2}) of the formfactor $F_W$
is determined by a
finite subtraction, which is needed for reproducing the correct value of
the trace anomaly fixed by a well-known low-energy theorem. In this way,
the result (\ref{eq2})
emerges as the difference between the anomaly value and a
sum rule for the imaginary part of the $F_W$ (see (\ref{eq5})).
On the other hand,
using the approximation invoked in \cite{b7} one gets an
``infrared''  interpretation
of the limit (\ref{eq2}) as an effect of the (would-be)
Goldstone bosons associated
with the $W$. Thus, the previous reasoning can be turned around: Using the
insight provided by the GB approximation \cite{b7} together with the sum rule
(\ref{eq5}), (\ref{eq6}) one arrives at an alternative (dispersive)
derivation of the $W$
boson contribution to the trace anomaly.

{\bf Acknowledgement}: We are grateful to J. Novotn\'{y} for
many useful discussions on the subject. One of us (J.H.) would
like to thank J.G.K\"{o}rner for a conversation which stimulated
this investigation. The work has been supported in part by the
research grants GACR -- 202/95/1460 and GAUK -- 166/95.

\end{document}